\documentclass[sigconf, nonacm]{acmart}
\usepackage{subcaption}
\usepackage{graphicx}
\usepackage{float} 
\AtBeginDocument{%
  }

\setcopyright{acmlicensed}
\copyrightyear{2025}
\acmYear{2025}
\acmDOI{XXXXXXX.XXXXXXX}
\acmConference[ICAIF ’25]{}{November 15–18, 2025}{Singapore}
\acmBooktitle{6th ACM International Conference on AI in Finance (ICAIF ’25), November 15–18, 2025, Singapore.}

\begin{document}

\title[Painting the market]{Painting the market: generative diffusion models for financial
limit order book simulation and forecasting}


\author{Alfred Backhouse}
\affiliation{%
 \institution{Department of Computer Science, University of Oxford}
 \country{United Kingdom}}

\author{Kang Li}
\affiliation{%
  \institution{Department of Statistics, University of Oxford}
  \country{United Kingdom}
}

\author{Jakob Foerster}
\affiliation{%
 \institution{Foerster Lab for AI Research, University of Oxford}
 \country{United Kingdom}}

\author{Anisoara Calinescu}
\affiliation{%
 \institution{Department of Computer Science, University of Oxford}
 \country{United Kingdom}}

\author{Stefan Zohren}
\affiliation{%
 \institution{Department of Engineering Science, University of Oxford}
 \country{United Kingdom}}


\begin{abstract}

Simulating limit order books (LOBs) has important applications across forecasting and backtesting for financial market data. However, deep generative models struggle in this context due to the high noise and complexity of the data. Previous work uses autoregressive models, although these experience error accumulation over longer-time sequences. We introduce a novel approach, converting LOB data into a structured image format, and applying diffusion models with inpainting to generate future LOB states. This method leverages spatio-temporal inductive biases in the order book and enables parallel generation of long sequences overcoming issues with error accumulation. We also publicly contribute to LOB-Bench, the industry benchmark for LOB generative models, to allow fair comparison between models using Level-2 and Level-3 order book data (with or without message level data respectively). We show that our model achieves state-of-the-art performance on LOB-Bench, despite using lower fidelity data as input. We also show that our method prioritises coherent global structures over local, high-fidelity details, providing significant improvements over existing methods on certain metrics. Overall, our method lays a strong foundation for future research into generative diffusion approaches to LOB modelling. 
\end{abstract}

\received{18 July 2025}

\maketitle

\section{Introduction}

Recently, deep generative models such as LLMs and diffusion models have shown progress across a range of domains \cite{brown2020language, ho2020denoising, stablediffusion}. Public progress in the financial domain is limited due to proprietary research and the complexity and noise of the data \cite{gould2013limit}. Accurately simulating financial markets has several useful applications \cite{li2025marsfinancialmarketsimulation}: forecasting future market moves, backtesting counterfactual impact to test trade ideas, creating realistic reinforcement learning environments for training agents, and identifying unexpected market outcomes. Therefore, generative modelling of financial data is a valuable and under-explored area. 

Limit order book (LOB) data details the full state of the order book --- the open buy and sell orders in the market --- giving available volume at every price level. Level-3 LOB data also includes every message sent to the exchange, such as new orders or cancellations. Previous work to model LOBs applies traditional methods such as linear models, convolutional neural networks or generative adversarial networks \cite{gould2013limit, zhang2019deeplob, coletta2021towards}. More recent work applies autoregressive sequence models \cite{nagy2023generative}. The standard industry benchmark, LOB-Bench \cite{nagy2025lobbench}, can be used to evaluate different generative LOB models across a range of metrics. 

Diffusion models have shown strong performance in image generation, due to their ability to leverage spatial inductive biases in the data. Previous work on LOBs \cite{zhang2019deeplob, bdlob, liu2023deep} treats order books as 2D images corresponding to the price levels and quantities over time, taking advantage of the hierarchical spatial features present in LOB data. We combine these ideas, converting LOB histories into our own, improved, image format, and applying diffusion models to provide an improved generative LOB model. We compare this against state-of-the-art models using the LOB-Bench benchmark, and find that our method achieves competitive performance under L1 loss and state-of-the-art performance under Wasserstein loss. Our method also provides faster inference of longer sequences. Overall, our novel diffusion-based method represents a promising new direction for generative LOB modelling and the full code will be made public if the paper is accepted. In summary, the main contributions of this paper are: 

\begin{itemize}
    \item An improved method for representing LOBs as images, better representing the structure of the data through a superior layout. 
    \item A novel approach to, and implementation of, generative LOB modelling using diffusion and inpainting on the new image representation.
    \item Public contribution to the industry benchmark for generative LOB models, LOB-Bench \cite{nagy2025lobbench}, allowing fair comparison between Level-2 (order book only) and Level-3 (order book and message) based methods. 
    \item A thorough evaluation of our image-based diffusion method, showing state-of-the-art performance on LOB-Bench, including examining its strengths and weaknesses through ablation studies on sample speed and robustness. 
\end{itemize}

\section{Background}

\subsection{Limit order books}

LOBs are the dominant microstructure for trading in modern electronic financial markets \cite{gould2013limit}. They operate in continuous time, allowing participants to submit buy and sell orders at any price and size. It is a continuous double auction, such that when a buy and sell order cross --- the bid, or buy price, is higher than the offer, or sell price --- the orders are executed for the maximum available size. Orders are matched based on price-time priority, meaning that orders with a more competitive price (higher for bids, lower for asks) are executed first, and for orders with the same price, the order which was submitted first has priority. Limit orders -- which specify a size and a price limit -- rest on the book until matched or cancelled, and the LOB is made up of outstanding limit orders. All messages (e.g. orders and cancellations) are anonymised and can be seen by all participants immediately as they are submitted. Level-2 LOB data consists only of LOB book states at every timestep while Level-3 data also includes every individual message. 

\subsection{Diffusion}

Generative models learn the underlying distribution of data and are able to sample and generate new, similar data. Diffusion models have shown particular aptitude in the image and video generation space \cite{betker2023improving, openai2024sora, liu2024sora}. A common framework for diffusion is denoising diffusion probabilistic models (DDPM) \cite{ho2020denoising} which we introduce here: the forward diffusion process adds Gaussian noise to data samples over $T$ timesteps, such that after the final step the sample is pure Gaussian noise. This allows us to produce partially noised samples which can be used during training. At each step, a small amount of noise is added: 
\[
q(\mathbf{x}_t \mid \mathbf{x}_{t-1}) = \mathcal{N}(\mathbf{x}_t; \sqrt{1 - \beta_t} \mathbf{x}_{t-1}, \beta_t \mathbf{I})
\]
with $\beta_t \in (0, 1)$ controlling the variance schedule.

Importantly, we can efficiently sample the noised data at any step using the closed-form marginal for $q(\mathbf{x}_t \mid \mathbf{x}_{0})$. During the reverse diffusion process, the goal of the model is to learn the reverse conditional, which is equivalent to predicting the image with one step of noise removed:
\[
p_\theta(\mathbf{x}_{t-1} \mid \mathbf{x}_t) = \mathcal{N}(\mathbf{x}_{t-1}; \boldsymbol{\mu}_\theta(\mathbf{x}_t, t), \sigma_t^2 \mathbf{I})
\]

Eventually, this allows the model to start with pure noise and incrementally generate new images. In practice, the model predicts the noise at time $t$, $\boldsymbol{\epsilon}_\theta$, rather than the previous image directly. For training, we sample an image from the training set, add a random amount of noise to it and predict the noise added at the final step. For inference, the model starts with a fully noised image $x_T$ and incrementally performs the full reverse diffusion process. This is done step by step by repeatedly predicting the noise and removing it until the model outputs $x_0$. While we have introduced the method in the context of images, the diffusion process could be applied across any continuous data.

\subsection{Inpainting}

For inpainting, the input to the model is an image with a noised region and the model learns to denoise the full image, using the unnoised regions as context to guide the reconstruction of the noised region \cite{lugmayr2022repaint, meng2021sdedit}. This mechanism aligns well with the LOB setting, where we have the LOB history in the natural image structure and want to fill in the future, which continues on smoothly from the image in the same format.

\section{Related work}

\subsection{Generative LOB modelling}

Traditional methods for generative LOB modelling use GANs \cite{coletta2021towards, coletta2023conditional, assefa2020generating} or LSTMs \cite{kong2025unlocking, lim2019enhancing}. More recently, the authors of \cite{nagy2023generative} use a token-level autoregressive model built on the S5 state space model architecture. The authors also provide LOB-Bench, an extensive suite of tools to benchmark generative LOB models \cite{nagy2025lobbench}. MARS \cite{li2025marsfinancialmarketsimulation} provides a market simulation engine and generative large market foundation model. Their model is built from multiple smaller models, both using autoregressive transformers. One encodes an overview of the behaviour of batches of the history, while the other generates fine grained individual orders, to provide a combination of global coherence and high fidelity. Unfortunately the code is not available and the authors have not tested their models on any benchmarks, so we could not directly compare with the MARS method. TRADES, a Transformer-based denoising diffusion probabilistic engine for LOB simulations \cite{berti2025TRADES}, combines Transformers with autoregressive time-series diffusion. 

\subsection{Deep learning for forecasting using images}

To explicitly represent the space-time structure of the LOB, LOB input data can be represented as images as in DeepLOB \cite{zhang2019deeplob} and its extensions \cite{bdlob, multideeplob, kisiel2022axial}. The authors transform the input data to a (100, 40) matrix, where each of the rows represent time (the 100 most recent book updates) and the columns represent the price and volume at each level. Specifically, an input image is defined as:
\begin{align*}
X &= [x_1, x_2, \cdots, x_t, \cdots, x_{100}]^T \in \mathbb{R}^{100 \times 40} \\
\text{where} \quad x_t &= [p_a^{(i)}(t), v_a^{(i)}(t), p_b^{(i)}(t), v_b^{(i)}(t)]_{i=1}^{n=10}
\end{align*}
where $p^{(i)}$ and $v^{(i)}$ are the prices and volumes at the $i$-th level of the book. They use the top 10 levels of the book, claiming that 80\% of the information is contained in the first level of the book and 90\% of orders are never filled so more than 10 levels adds unnecessary noise \cite{zhang2019deeplob, Cao2008}. Due to the mixing of asks, bids, prices and volumes, the image layout requires certain handcrafted convolutional filters to process it. We introduce improvements to the layout later.

Further work \cite{liu2023deep} provides an alternative image format for a multi-asset setting. However, to reduce the input dimension they ignore the full LOB book, using just mid price returns, with each row representing a timestep (as before), and each column representing a different asset's mid price return since the previous timestep. Using a deep inception network, they output the predicted optimal portfolio construction at each timestep. 

The authors of \cite{li2025marsfinancialmarketsimulation} use the order batch model to model longer term dependencies, as discussed in the generative section above, encoding order batches as images in a unique way to provide a summary of the number of bid, ask and cancel orders at each price and volume level. However, as this method serves to compress approximate longer term information into a compact image, which is then further encoded into a single token using a visual model, it is not useful for high-fidelity LOB generation. 

Over the background and related work sections we have covered existing LOBs, diffusion methods, generative LOB models, and previous work to model LOBs as images. Combining these concepts together provides a clear and novel direction for our work, which we discuss in the next section. 

\section{Methodology}

By leveraging the inherent spatio-temporal structure of LOBs using an image format and applying diffusion models, we can create an improved paradigm for generative LOB models. Over this section, we give more detailed theoretical motivation for the approach and explain the implementation details. We also explain the contributions to the LOB-Bench benchmark to allow fair comparison between Level-2 and Level-3 based models. 

\subsection{Theoretical grounding and motivation}

High frequency financial data is extremely noisy due to the large number of participants, high cancellation rate, asynchronous order flow and market microstructure effects. This makes extracting signal from raw data hard, especially without compressed structure or inductive biases. 

We claim that an image-based approach with a convolutional model has strong theoretical advantages \cite{726791}:

\textbf{Translation invariance} --- convolutions are naturally translation invariant. The order book will constantly be shifted along the temporal axis as we move through time, and we want the model to be invariant to these shifts. Similarly, shifts in price levels are common due to entire levels being cancelled or filled. \textbf{Localised feature extraction} --- convolutional model is suited to detecting and compressing local patterns in price and time. \textbf{Weight sharing} --- a convolutional model uses shared weights across price level and time. This makes it more parameter efficient and less prone to overfitting, which is particularly relevant due to the high noise in the data. \textbf{Inductive bias towards stationarity} --- local structures and behaviours are likely to repeat across price and time. \textbf{Hierarchical feature extraction} --- many of the features to extract are composed of simple localised events. CNNs excel at aggregating hierarchical features into more complex features through layers of filters using the composition of simple events \cite{sermanet2014overfeat}. \textbf{Interpretability} --- feature maps in CNNs can be more interpretable and can attribute final outputs to specific price-time regions through saliency maps or class activation maximisation \cite{simonyan2013deep}. This is particularly relevant in the financial setting, where human-machine interaction and understanding is necessary \cite{quinn2023explaining}.

We also claim that an inpainting diffusion model provides significant benefits over an autoregressive approach:

\textbf{Convolutional backbone} --- many diffusion methods use a convolutional backbone \cite{stablediffusion, openai2024sora}, which is useful for all the reasons above. \textbf{State-of-the-art in vision} --- assuming that the image-based approach is sound, then the diffusion approach is a natural choice. \textbf{Structured noise} --- noise is inherent in financial markets \cite{gould2013limit} and diffusion models are naturally able to handle continuous data and small perturbations in a stable fashion without overfitting.  \textbf{Parallel generation of future timesteps} ---  diffusion models predict the whole output simultaneously. This possibly leads to faster inference for longer-time sequences  \cite{ho2020denoising}, although is counterbalanced by the large number of inference steps required for iterative denoising.  \textbf{Diversity of generated outputs} --- diffusion models are especially strong at providing a diverse range of generated outputs \cite{imagen}. This is particularly relevant in the financial setting due to heavy distributional tails and diverse real data distribution. \textbf{Parallel generation of possible outputs} --- diffusion models are able to generate a range of possible outputs in parallel by varying the noise seed across the batch. \textbf{Mitigating compound error} --- autoregressive models progressively predict the next token during generation. The prediction error at each step accumulates and propagates to subsequent steps. This phenomenon is called the compound error. However, diffusion models generate the entire sequence at once through a denoising process, with no step-by-step dependencies --- the generation errors of each token are independent of each other, avoiding the step-by-step accumulation and propagation of errors. Compared to autoregressive models, the compound error of diffusion models is smaller \cite{li2022diffusion}.

This provides a strong motivation for using an image-based approach and a diffusion model with a convolutional backbone. We discuss the details of our approach over the upcoming sections. 

\subsection{LOB data as images}

Inspired by previous work \cite{zhang2019deeplob}, we use an image-based representation of the LOB. We also improve the layout, to provide more meaningful temporal and spatial layout, as follows: given a long data stream of LOB states, we sample a fixed-length window of $T$ states and take the top $n$ levels of the book.

\begin{figure}[t]
    \centering
    \includegraphics[width=0.7\linewidth]{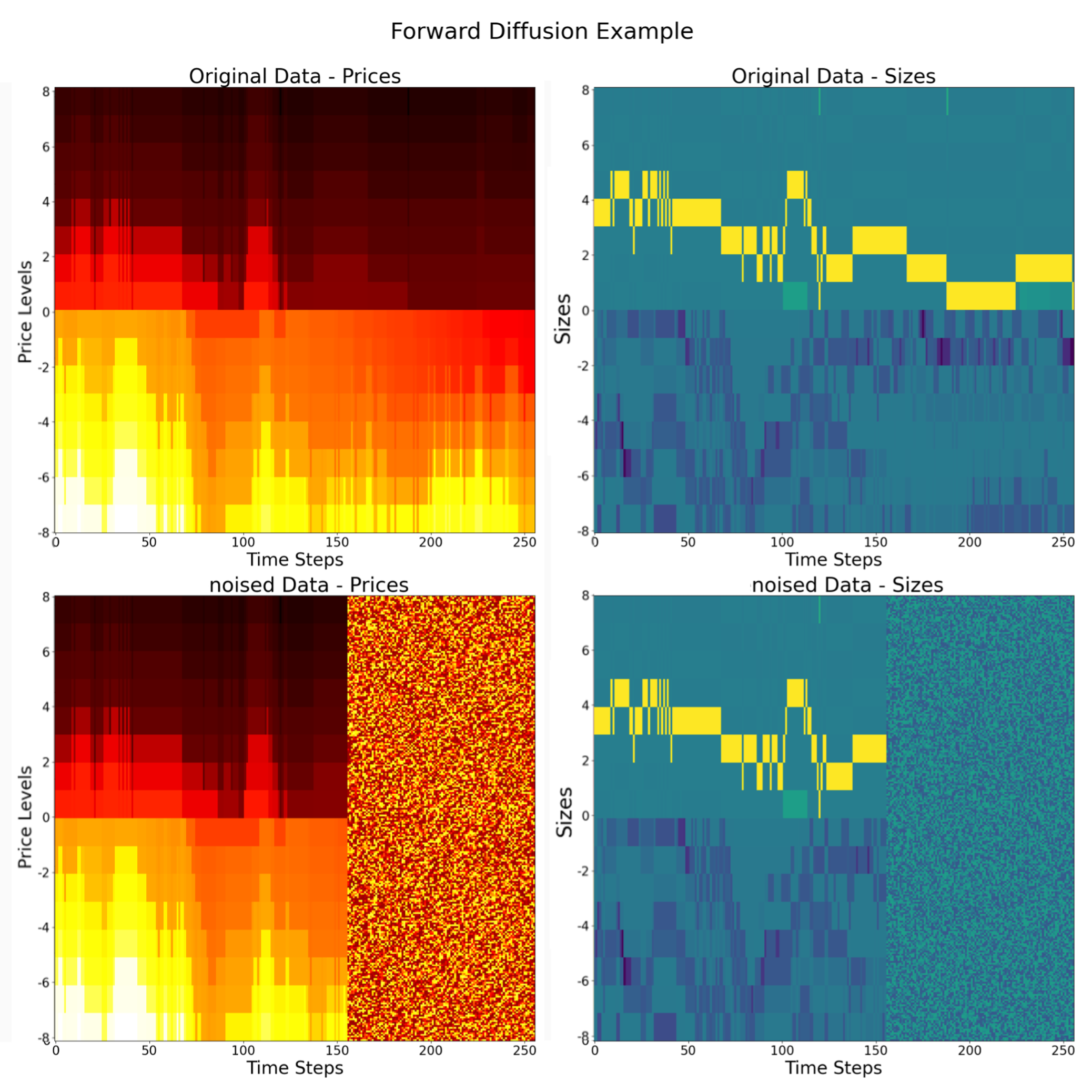}
    \caption{An example of the forward diffusion process. The top row shows the original data in the image format, with prices on the left and sizes on the right. Each column represents a snapshot of the order book at each timestep. The bottom row shows a fully noised input to the model, which would be fed in at inference time. The first 156 timesteps represent the past, and are unnoised. The next 100 timesteps are pure noise representing the future for the model to predict.}
    \label{fig:forward_diffusion}
\end{figure}

We construct a 2-channel image $X \in \mathbb{R}^{2n \times T \times 2}$, with a visual example shown in Figure \ref{fig:forward_diffusion}, where:
\begin{itemize}
    \item Each column represents an order book state. Moving along the width dimension from column 1 to $T$ moves forward in time.
    \item Each row represents a price level. Moving along the height dimension from 1 to $2n$ increases in price, i.e. starting from the lowest, least competitive bid price, we move towards the highest, most competitive bid price. The lowest, most competitive ask price is directly above the best bid, and the highest, least competitive ask price is at the top of the image. 
    \item The first channel contains prices, and the second channel contains sizes.
\end{itemize}

To normalise, let $p_a^{(i)}(t)$ and $p_b^{(i)}(t)$ denote the ask and bid prices at level $i$ and timestep $t$, and $v_a^{(i)}(t)$, $v_b^{(i)}(t)$ the corresponding sizes. At the initial timestep $t_0$ of each window, we compute the mid price, $text_{mid}(t_0) = (p_a^{(1)}(t_0) + p_b^{(1)}(t_0))/2$.
All subsequent prices in the window are centred by subtracting $\text{mid}(t_0)$. Ask sizes are multiplied by $-1$ and then sizes are assumed to be zero-mean. Normalisation is completed by dividing by a rolling standard deviation of the prices and sizes. 
Following \cite{nagy2023generative}, we clip the data to the 95 percentile to avoid extreme anomalies. 


Given the excess amount of training data, we step each training example forward by the full history length, ensuring that each training example contains entirely fresh data.

For passing the data into our model, we follow standard inpainting procedures and pass in three stacked images:  the first is the entire original image with noise added to both the history and future parts, containing a price and a size channel. The second is the unnoised history, with the future fully masked out (set to zeros). This also contains a price and a size channel. The third is the inpainting mask, with zeros in the first 64 timesteps and ones in the second 64 timesteps. Overall this creates a (batch-size, $T$, $n$, 5) input image. 

\subsection{Model}

We use an existing Hugging Face implementation of the unconditional UNet model \cite{huggingface_diffusers} --- a well-documented standard framework widely used in recent state-of-the-art diffusion research \cite{stablediffusion, podell2023sdxl, couairon2022diffedit}.  UNets use a convolutional backbone to enforce strong inductive biases. Unlike many modern implementations \cite{stablediffusion}, our model operates directly in the image space, rather than a latent space, to allow easy alignment of the historical mask.

Figure \ref{fig:forward_diffusion} shows an example of a fully noised input to the model: the first 156 timesteps represent the history, and the remaining 100 timesteps represent the fully noised future for the model to generate. Prices and volumes are fully normalised and displayed as relative values through a heatmap. Original scaled values can be recovered after prediction.

Standard UNet architectures require a square input and output images. Therefore, we pad the level dimension by repeating each level so that the model takes an input of shape $(T, T, 5)$ and an output of shape $(T, T, 2)$ which can be undone after prediction. The architecture uses 6 downsampling and 6 upsampling layers, with the number of filters increasing with depth in the sequence $(128, 128, 256, 256, 512, 512)$. Each downsampling block consists of two convolutional layers followed by normalisation, ReLU activation and pooling (to perform the downsampling). The fifth block introduces self-attention to capture long-range dependencies at the cost of a higher parameter count. The upsampling blocks mirror these and include attention in the second layer.



\begin{figure}[t]
    \centering
    \includegraphics[width=1\linewidth]{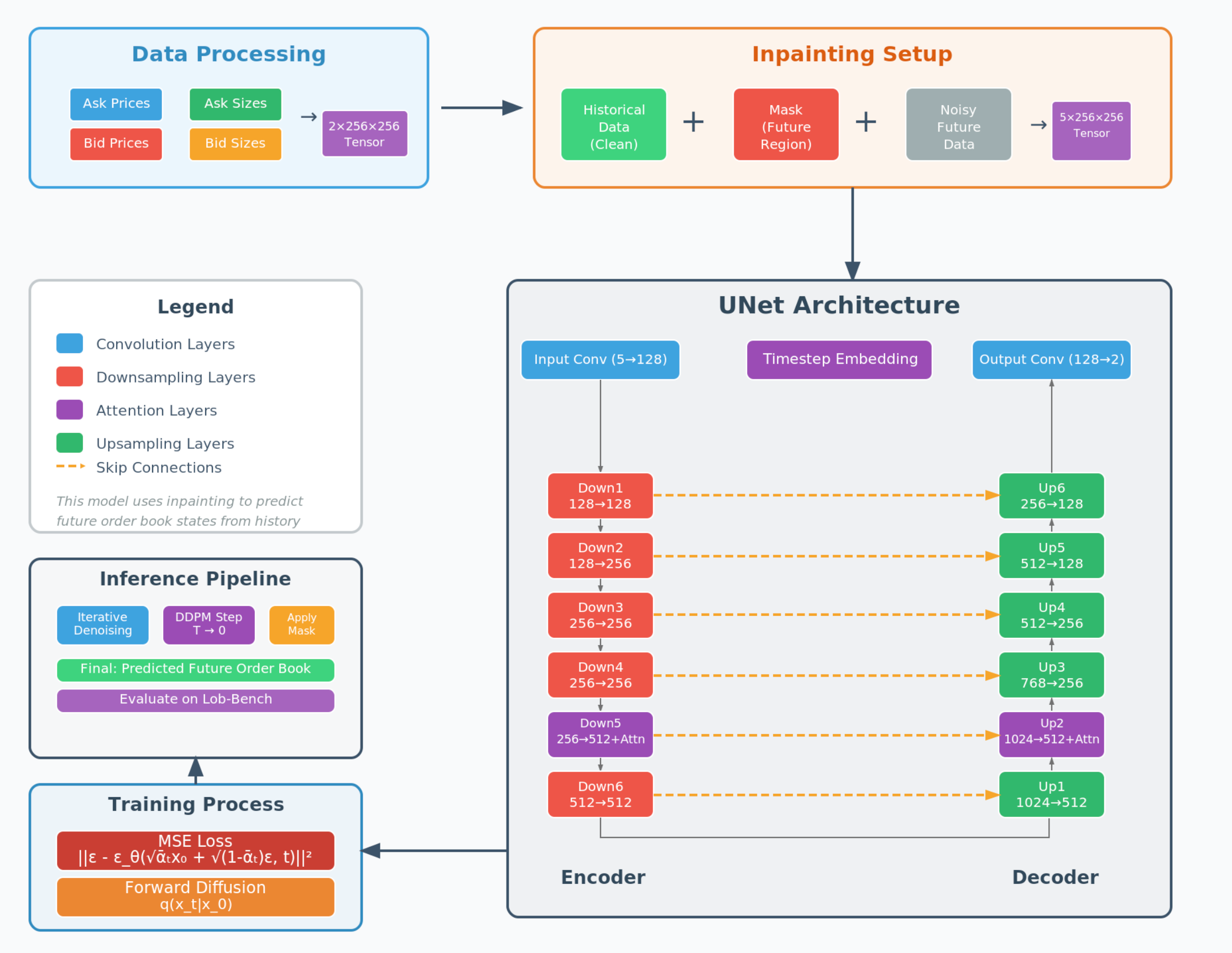}
    \caption{Summary of the end-to-end pipeline for our generative diffusion model with inpainting.}
    \label{fig:overall_method}
\end{figure}

\section{Experiments}

\subsection{Data and experimental setup}

We focus on the same examples of a small-tick (Alphabet, GOOG) and large-tick (Intel, INTC) stocks used in LOB-Bench \cite{nagy2025lobbench}, using the LOBSTER dataset \cite{Huang2011}. Small-tick stocks contain sparser order books and more volatile spreads than large-tick stocks. We restrict the data to regular trading hours and use 102 days of training data (1 July 2022 to 11 November 2022), 12 days of validation (28 November 2022 to 13 December 2022) and 12 days of test data (14 December 2022 to 30 December 2022).
Unlike \emph{LOBS5}, our model cannot use Level-3 message data and considers only Level-2 order book data. This means that we have strictly less data and information available. Consistent with \cite{nagy2025lobbench}, we use a classical baseline model from Cont et al.  \cite{cont2010stochastic}, and existing generative models \emph{Coletta}  \cite{coletta2022learning}, \emph{RWKV-6} \cite{peng2023rwkv, peng2024eagle} and \emph{LOBS5} \cite{nagy2025lobbench}. 

\subsection{Model training and qualitative examples}

We use a single Nvidia A40 GPU for training, and train separate models on the INTC and GOOG data, as per \cite{nagy2023generative}. We train the model on a single pass through the data (epoch) with no repetition, since there is an excess of high-quality recent order book data. Training takes 18 hours for INTC and 34 hours for GOOG, with the difference due to the higher rate of order book updates for GOOG which gives a larger training set within the same time window.

Training time increases quadratically with the height of the image. As a result we limit ourselves to a history of 156, and a prediction length of 100. This matches the original \emph{LOBS5} which has a prediction window of 100, while maintaining a power of 2 for the resolution for computational efficiency.

Under these constraints, we train our models and provide qualitative examples and stylistic analysis. Our model performs significantly better on GOOG than INTC, both in perceived visual quality and in comparison to other models later on. We believe this is due to the large ticks in INTC, which discourages frequent spread-crossing and decreases high frequency volatility \cite{eisler2012price, zhang2025clusterlob}. As a result, price changes are infrequent and signal is sparse in the data, reducing the amount of useful information in short term order book histories and limiting the model's predictive power. Methods to overcome this limitation are discussed in the conclusion. 

\begin{figure}[t]
    \centering
    \includegraphics[width=.7\linewidth]{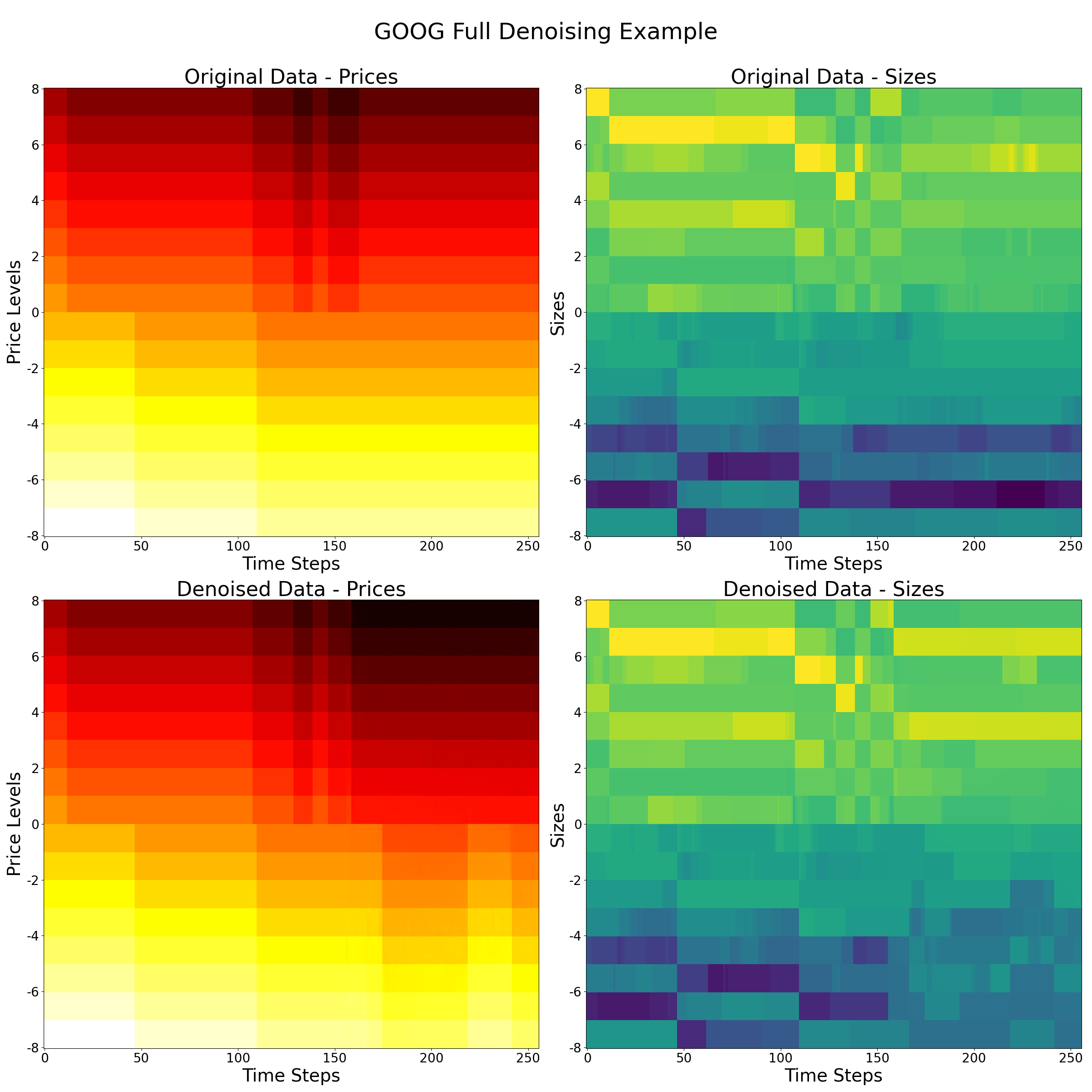}
    \caption{A typical example generated by our model on GOOG data. The top row shows the ground truth; the bottom row shows the generated example. The final 100 timesteps (after timestep 156) are fully noised and generated by the model.}
    \label{fig:goog_model_examples}
\end{figure}

\begin{figure}[t]
    \centering
    \includegraphics[width=.7\linewidth]{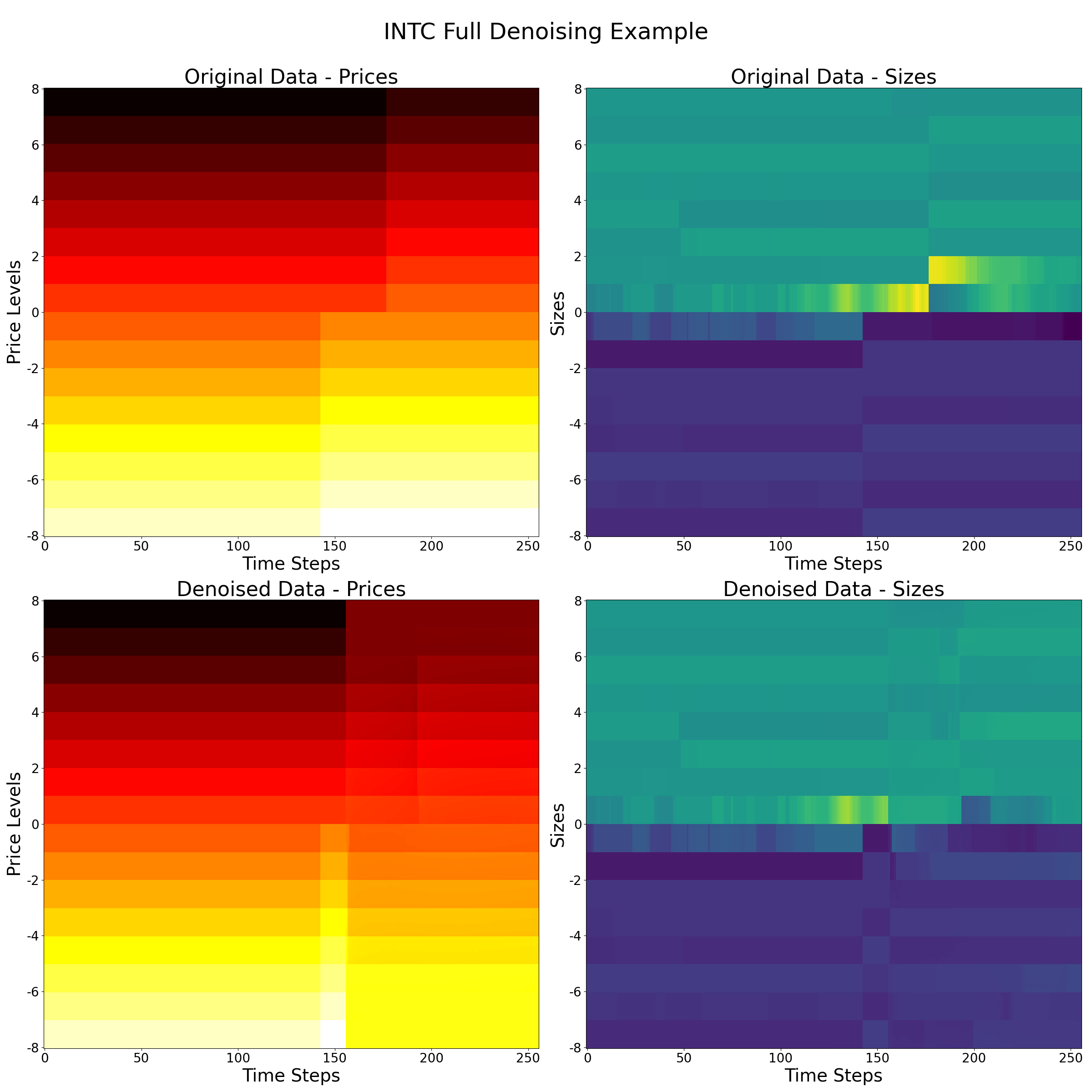}
    \caption{Example generated by our model on INTC data. The top row shows the ground truth, the bottom row shows the generated prediction. A sharp division at timestep 156 is visible, where the output collapses to the mode.}
    \label{fig:intc_model_examples}
\end{figure}

The model can be seen to struggle significantly on INTC data, as shown in Figure \ref{fig:intc_model_examples}. The model displays mode collapse on the prices, where after timestep 156, the output roughly collapses to a modal output and a sharp division can be seen. These limitations in the INTC model could potentially be overcome with better data formatting and compression or larger models but leave this to future work. Therefore, for this work we focus on the GOOG model as our flagship model for evaluation. 

Following previous work \cite{nagy2023generative, nagy2025lobbench}, we compared the overall distributions of the real and generated data for the GOOG model.  The model is able to capture the broad target distribution across key measures such as spread, order book imbalance and volumes. There is a notable underestimation of high-volume orders - this could be partly due to the clipping of high-volume events in the training data, or purely a result of over-smoothing of volume predictions.

Overall, these results show that both models are able to capture the high-level statistical properties of the order book. The GOOG model is able to produce insightful and visually plausible outputs, while the INTC model displays mode collapse, suggesting a failure to extract valuable insight from the history, likely due to the sparse signal in the data. We now turn to LOB-Bench to provide quantitative comparisons with the existing models. 

\subsection{Comparison with generative message methods}

We compare our model to existing methods from LOB-Bench, using the same train and test set. Despite the fact that our model only has access to Level-2 order book data, compared to the richer Level-3 message and order data used for \emph{RWKV} and \emph{LOBS5}, our model performs comparably on GOOG across most metrics. 

\begin{figure}[ht]
    \centering
    \begin{subfigure}[b]{0.35\textwidth}
        \includegraphics[width=\textwidth]{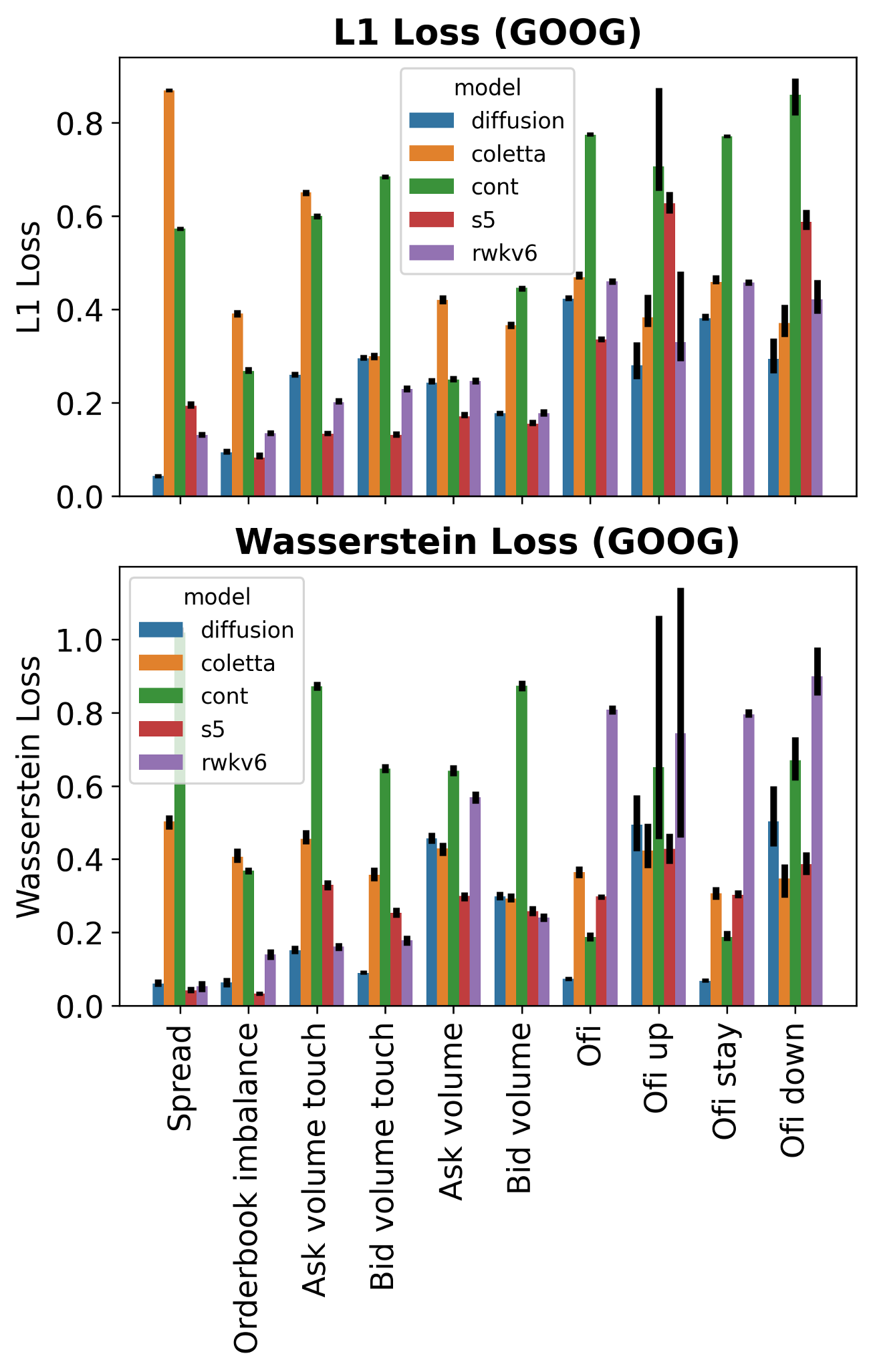}
    \end{subfigure}
    \caption{Model comparison on GOOG, showing comparable, or superior, performance to \emph{LOBS5} across the different metrics.}
    \label{fig:comparison_bars_GOOG}
\end{figure}

\begin{figure}[ht]
    \centering
    \includegraphics[width=\linewidth]{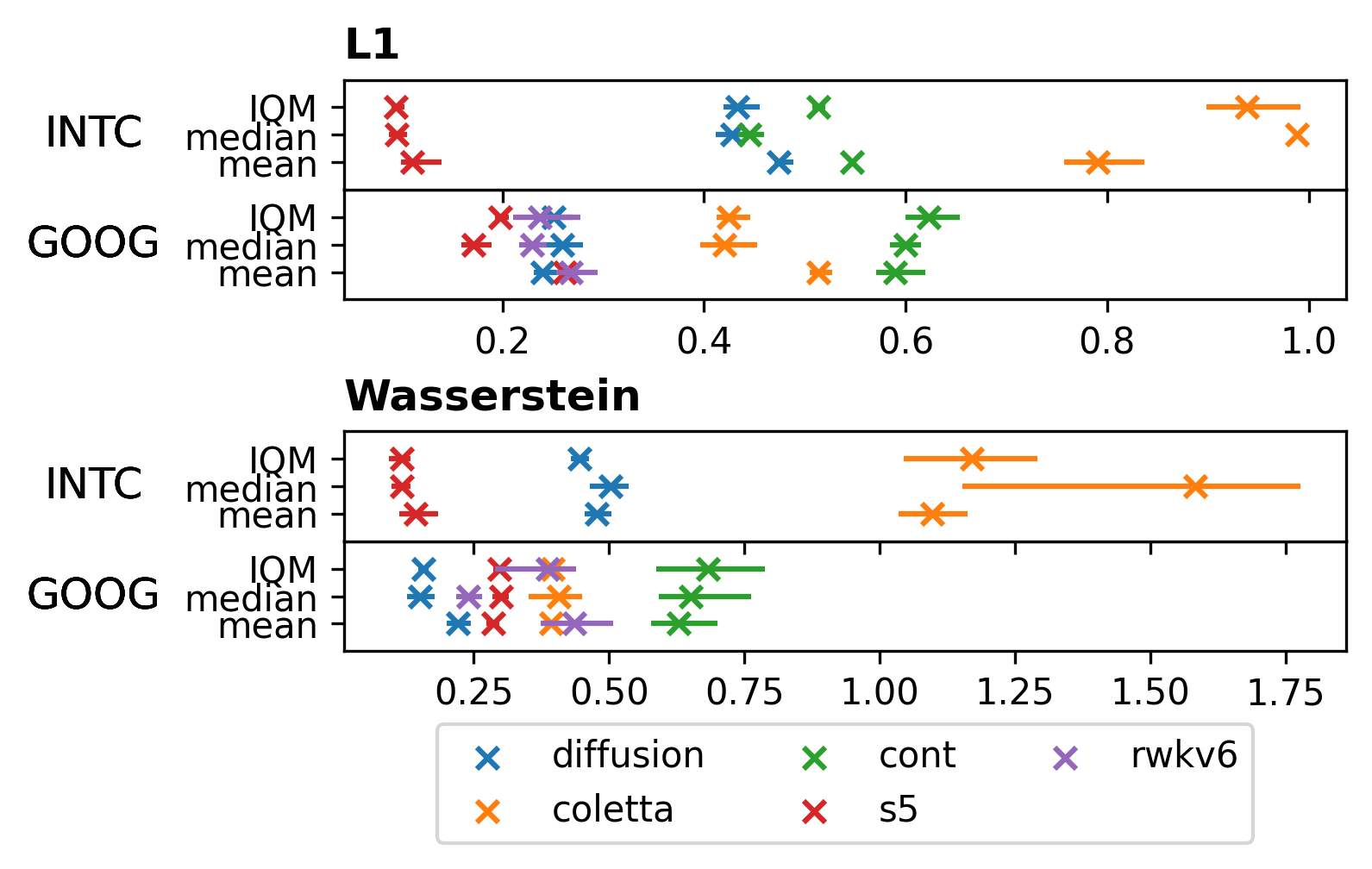}
    \caption{Summary statistics for all the models on GOOG and INTC. For INTC, \emph{LOBS5} is dominant and outperforms our model easily. For GOOG, our model is competitive under L1 loss and achieves state-of-the-art performance under Wasserstein loss.}
    \label{fig:summary_stats_full}
\end{figure}

Figure \ref{fig:comparison_bars_GOOG} show that the model achieves strong results on GOOG, outperforming the state-of-the-art \emph{LOBS5} model on spread under L1 loss and achieving state-of-the-art performance across several metrics under Wasserstein. Overall, Figure \ref{fig:summary_stats_full} shows that our diffusion model beats all models except \emph{LOBS5} when evaluated on L1 loss, and achieves state-of-the-art on Wasserstein loss overall. 

This can be understood by considering that L1 loss captures average absolute errors and is sensitive to localised deviations, while Wasserstein loss captures the distance between entire distributions and is sensitive to global shape and structure. This aligns with some of the distributions seen in the stylised facts for our model on GOOG, where the generated data looks like a smoothed out version of the real data. This also aligns with the example images, where the volumes could look smoothed. Furthermore, conventional wisdom says that diffusion models are especially good at modelling a global distribution but may smooth over small details and create blurry outputs. 

\begin{figure}[ht]
    \centering
    \begin{subfigure}[b]{0.35\textwidth}
        \includegraphics[width=\textwidth]{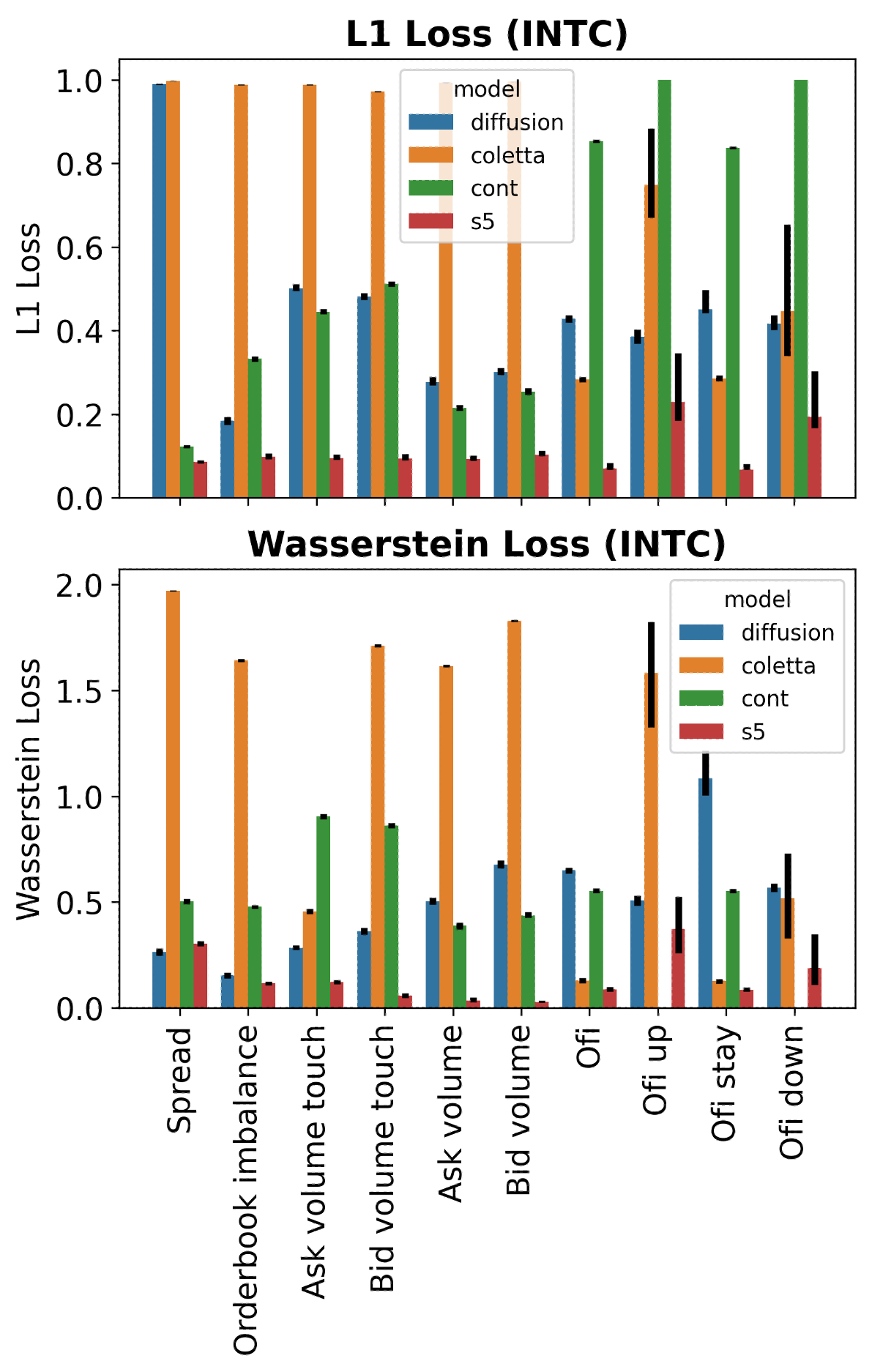}
    \end{subfigure}
    \caption{Model comparison on INTC, showing weak performance on most metrics, although still superior to \emph{Coletta}.}
    \label{fig:comparison_bars_INTC}
\end{figure}

In contrast, and as discussed previously, model performance on INTC is significantly worse than the benchmark models. A comparison is shown in Figure \ref{fig:comparison_bars_INTC}. This aligns with the mode collapse and model failure seen in the qualitative examples and reflects the difficulty of learning from sparse Level-2 data. 

Overall, these results demonstrate that our method can achieve state-of-the-art performance on GOOG, despite being trained and tested on less informative data. Therefore, the method provides a strong alternative to \emph{LOBS5}, especially when Level-3 message data is not available.

\subsection{Sample time ablation}

\begin{figure}[ht]
    \centering
    \begin{subfigure}[b]{0.35\textwidth}
        \includegraphics[width=\textwidth]{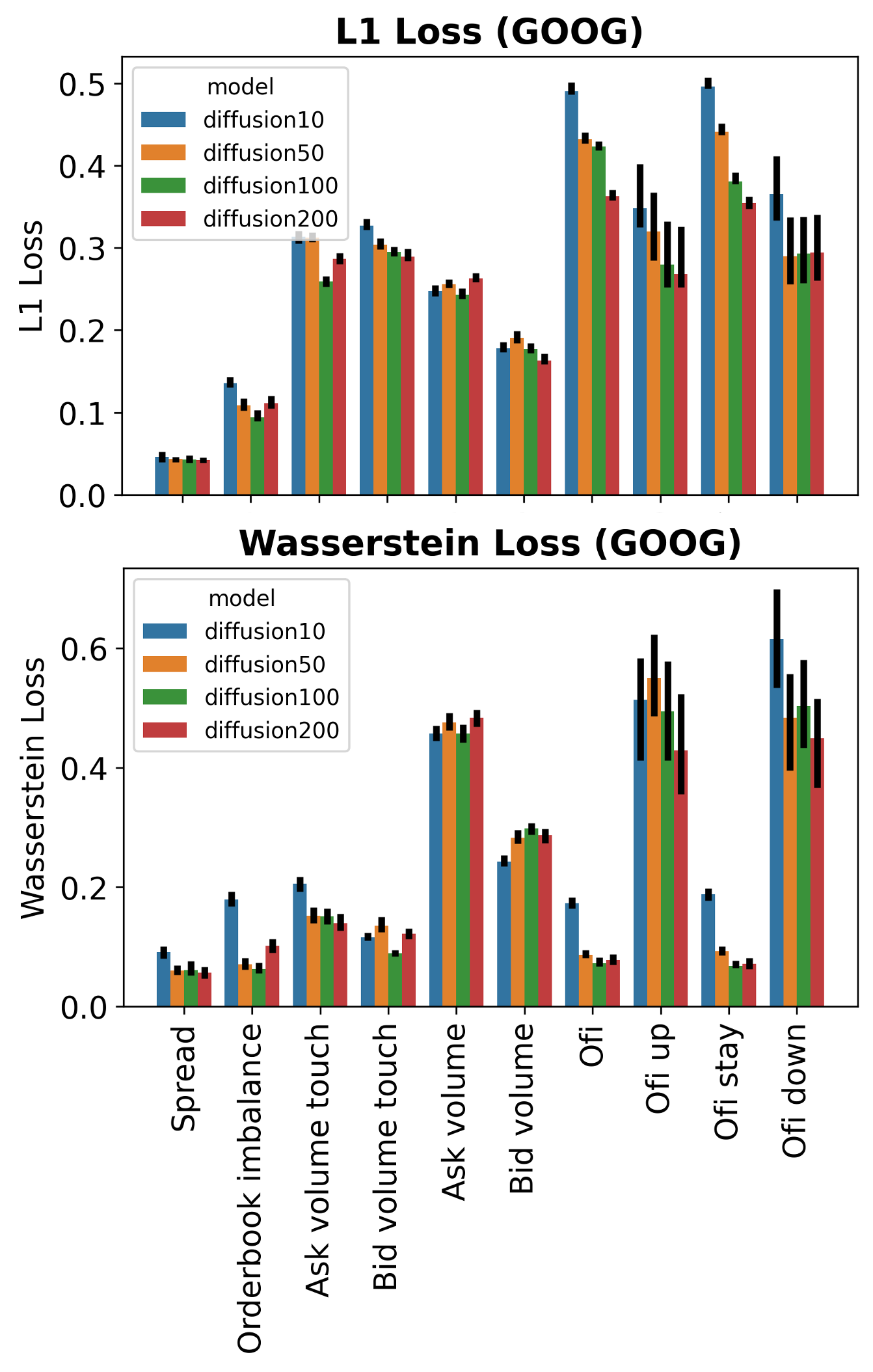}
    \end{subfigure}
    \caption{Sample time ablation results for reduced inference steps on GOOG. This shows that increased inference steps generally leads to a very small improvement in performance.}
    \label{fig:sample_time_bars_GOOG}
\end{figure}

Fast inference is essential for real world applications of these models. This is true for real time forecasting and trading, where the outputs are only useful if they are predicted before the events unfold, which could be a matter of milliseconds. It is also true of offline applications (backtesting, simulating counterfactuals, training RL agents) where we need to process large amounts of data and inference speed could be a bottleneck. Diffusion models naturally lead themselves to fast parallel generation as they predict the entire image at once, as opposed to step by step with autoregressive models. However, standard diffusion inference under our training regime takes 1000 steps to generate the full image, each requiring a forward pass through the model. In the search for faster inference while maintaining quick parallel generation, we explore the effect of reducing the number of sampling steps during test inference while keeping the model setup and training fixed. Specifically, using the same pre-trained model weights (trained on 1000 inference steps), we test it with 10, 50, 100 and 200 inference steps on INTC and GOOG. Figure \ref{fig:sample_time_bars_GOOG} shows that increasing the number of inference steps generally leads to better performance across the board, but the difference is small, especially between 100 and 200 steps, illustrating the diminishing returns of adding more steps. For many of the metrics, the scores are within the confidence intervals. For some of the metrics such as Order Flow Imbalance (OFI), increasing steps more clearly leads to improved performance. Similar patterns emerge for both GOOG and INTC. 

The differences are larger under L1 than Wasserstein loss for a similar reason to before - a low number of sampling steps is still able to capture the rough overall distribution, but lacks the fidelity prioritised by L1. Depending on the application, sacrificing fidelity for fast inference, and remembering that increasing sample steps takes linear time so 10 steps is 100 times faster than 1000 steps, could be worthwhile. These results highlight a key strength of our approach: diffusion models are able to generate entire LOB sequences in parallel, perhaps even with very few sampling steps, leading to high throughput for longer prediction horizons. 

\subsection{Generalisability or robustness ablation}

For real world applications, it would be useful if the model generalised well and was robust to regime changes. Regime changes mean entering a period with a distributional shift such as increased volatility. It would also be useful if we could apply a single model to all stocks, or if our model generalised to new, unseen stocks. 

\begin{figure}[ht]
    \centering
    \begin{subfigure}[b]{0.35\textwidth}
        \includegraphics[width=\textwidth]{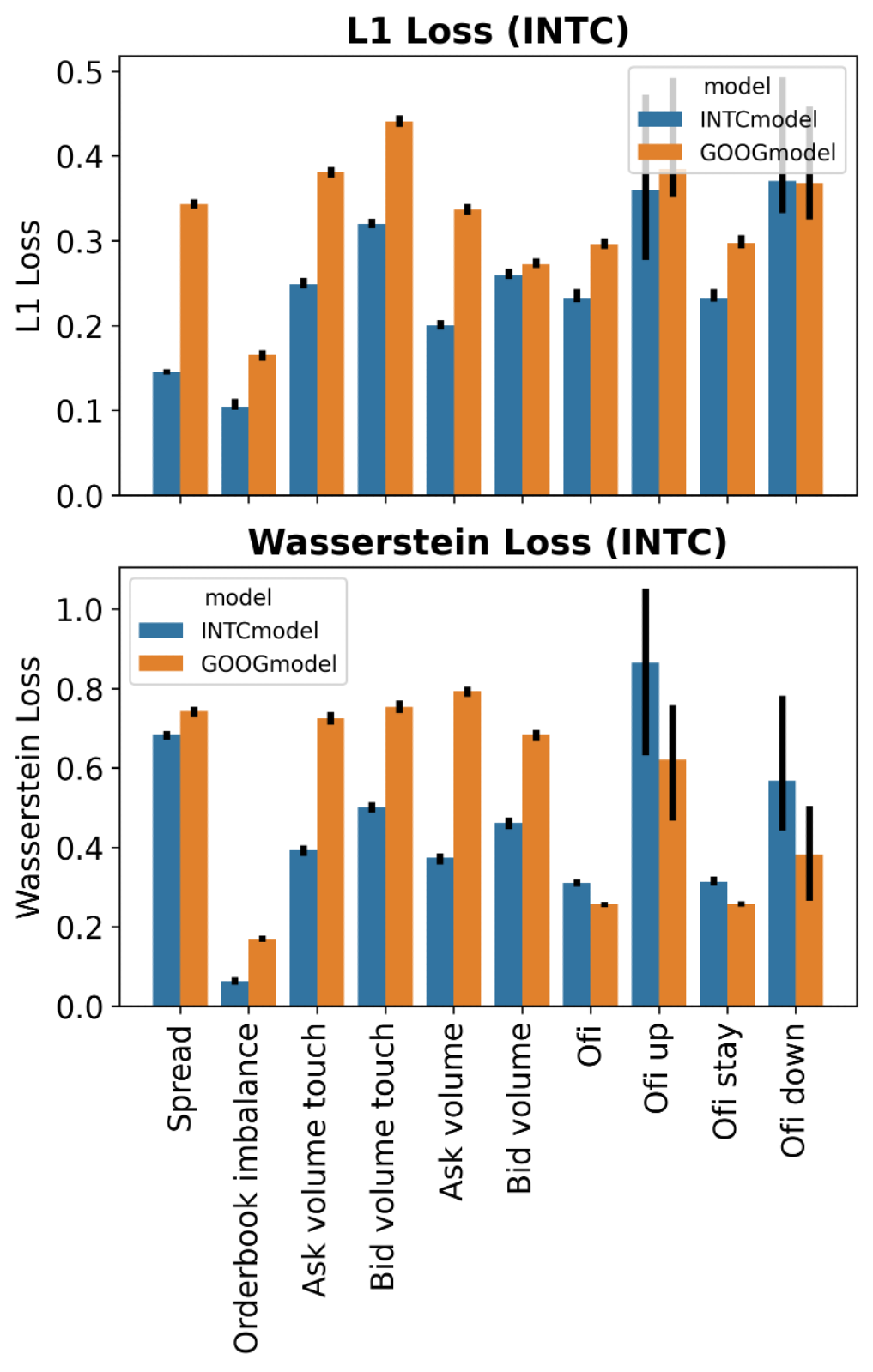}
    \end{subfigure}
    \caption{Robustness test on INTC data, showing the GOOG model's weak generalisability.}
    \label{fig:robustness_bars_INTC}
\end{figure}

To test this, we test the models trained exclusively on GOOG or INTC data on other data. Somewhat predictably, the INTC model, which performed poorly even on the INTC test data, performs significantly worse than the GOOG model in all areas. This is unsurprising. More interestingly, Figure \ref{fig:robustness_bars_INTC} shows the different models' performance on INTC test data. Here, even though the INTC model exhibits mode collapse, it still outperforms the GOOG model in all areas on L1 loss (which favours local precision and should penalise mode collapse more heavily). The results are similar for the Wasserstein loss, although the GOOG model outperforms on OFI metrics. This shows poor generalisability for GOOG. 


Overall, the results from the various experiments demonstrate that our diffusion based method is able to model complex order book dynamics and compete with the current state-of-the-art in the GOOG setting. Performance on INTC is initially weaker, although the model still models global distributional properties and enables fast, parallel inference. Ablation studies show that sample speed can be increased with minimal performance degradation however the model struggles with robustness and generalisability across different stocks and domains. These findings show the strength of our approach and lay the groundwork for further exploration. 

\section{Conclusion}

\subsection{Contributions}

We provide a number of significant contributions to the LOB modelling field. We publicly contribute to LOB-Bench, enabling a direct comparison of generative LOB models with Level-2 and Level-3 data. We introduce an improved image representation of LOBs and an image-based diffusion framework for generative LOB modelling, leveraging the spatio-temporal structure of the data. This is a novel approach to the problem, grounded in theoretical considerations and empirical performance. Despite only using Level-2 data, our model performs comparably with the state-of-the-art Level-3 models, and we examine its sample efficiency and robustness through rigorous ablations. Our work shows that image-based diffusion models are a competitive approach to generative LOB modelling, particularly when Level-3 data is not available or we want to prioritise distributional similarity over high local fidelity prediction.

\subsection{Future work}

Future work could perform diffusion in the latent space, as is common in other diffusion implementations, enabling compression of the input for a larger context window and reducing redundant information for sparse tickers. This could help our INTC model in particular. Manual compression, such as only taking the order book updates every few timesteps, could provide an easier alternative solution. 
The method could also be tested using a conditional, rather than inpainting based method. While we may lose some of the interpretability, inductive bias and continuity of the inpainting method, we may gain efficiency as the history is a conditional component and we only need to run explicit denoising on the prediction window (currently we run denoising on the history and prediction, and discard the history part). Future work could also implement other small improvements to the diffusion method such as DDIM, progressive distillation, or thorough hyperparameter sweeps. 


\subsection{Final thoughts}

Our method achieves state-of-the-art performance on the industry benchmark, LOB-Bench. While it currently faces challenges with large-tick stocks and generalisability, it introduces a promising new paradigm for generative financial modelling. With more data and further refinements, diffusion-based models could play a key role in the future of LOB simulation and forecasting.



\bibliographystyle{ACM-Reference-Format}
\bibliography{sample-base}

\end{document}